\documentclass[12pt]{article}
\usepackage{epsfig}

\newcommand{\np}[3]{Nucl.\ Phys.\ {\bf B#1} (19#2) #3}
\newcommand{\prd}[3]{Phys.\ Rev.\ {\bf D#1} (19#2) #3}

\newcommand{\eps}{\epsilon}

\def\simgt{\rlap{\lower 3.5 pt \hbox{$\mathchar \sim$}} \raise 1pt \hbox {$>$}}
\def\simlt{\rlap{\lower 3.5 pt \hbox{$\mathchar \sim$}} \raise 1pt \hbox {$<$}}

\newcommand{\beq}{\begin{equation}}
\newcommand{\eeq}{\end{equation}}
\newcommand{\bea}{\begin{eqnarray}}
\newcommand{\eea}{\end{eqnarray}}

\newcommand{\go}{\rightarrow}

\newcommand{\tb}{\mbox{tg$\beta$}}

\newcommand{\gsim}{\raisebox{-0.13cm}{~\shortstack{$>$ \\[-0.07cm] $\sim$}}~}
\def\ep{\epsilon}
\def\msb{\overline{\rm MS}}

\catcode`\@=11
\def\section{\@startsection{section}{1}{\z@}{3.5ex plus 1ex minus .2ex}
{2.3ex plus .2ex}{\large\bf}}
\def\thesection{\arabic{section}.}

\def\appendix{\setcounter{section}{0}
 \def\thesection{Appendix \Alph{section}:}
 \def\theequation{\Alph{section}.\arabic{equation}}}
\def\@citex[#1]#2{\if@filesw\immediate\write\@auxout{\string\citation{#2}}\fi
  \def\@citea{}\@cite{\@for\@citeb:=#2\do
    {\@citea\def\@citea{,\penalty\@m}\@ifundefined
       {b@\@citeb}{{\bf ?}\@warning
       {Citation `\@citeb' on page \thepage \space undefined}}%
\hbox{\csname b@\@citeb\endcsname}}}{#1}}

\def\citer{\@ifnextchar [{\@tempswatrue\@citexr}{\@tempswafalse\@citexr[]}}
\def\@citexr[#1]#2{\if@filesw\immediate\write\@auxout{\string\citation{#2}}\fi
  \def\@citea{}\@cite{\@for\@citeb:=#2\do
    {\@citea\def\@citea{--\penalty\@m}\@ifundefined
       {b@\@citeb}{{\bf ?}\@warning
       {Citation `\@citeb' on page \thepage \space undefined}}%
\hbox{\csname b@\@citeb\endcsname}}}{#1}}
\relax
\begin{document}
% ----------------------------------------------------------------------------
\thispagestyle{empty}
\begin{flushright}
ITP-SB-97-26\\
May 1996\\
hep-ph/yymmxxx
\end{flushright}
\vskip 1.5cm
\begin{center}{\bf\Large\sc Sudakov Effects in Higgs Boson Production}
\vglue .3cm
{\bf\Large\sc at the LHC}
\vglue 1.2cm
\begin{sc}
Eric Laenen\\
\vglue 0.5cm
\end{sc}
{\it Institute for Theoretical Physics \\
State University of New York at Stony Brook, NY 11794, USA}
\vglue 0.5cm
{\it Talk given at 32nd Rencontres de Moriond: QCD and High-Energy 
Hadronic Interactions, Les Arcs, France, March 1997}
\vglue 1.2cm
\end{center}

\begin{abstract}
We discuss the resummation of Sudakov effects in a cross section
from the viewpoint of its underlying factorization 
near the edge of phase space.
We perform the resummation of Sudakov threshold logarithms
in Higgs (Standard Model and Minimal Supersymmetric SM) 
production at the LHC, using an evolution
equation in the Higgs mass that is derived from 
this factorization. We extend the class of universal
large terms that is resummed 
to include additional universal contributions, which, when included,
help to reproduce the exact result to within a few
percent for the full allowed range of Higgs boson masses
in the SM and MSSM. Using the analytic resummed formula
as a generating functional for approximate perturbation theory, we
show results for next-to-next-to-leading order corrections
in Higgs production, and find they are potentially sizable.
\end{abstract}

\vfill

\section{Sudakov Factorization and Resummation}

The intimate connection between resummation of large logarithms 
in amplitudes in quantum field theory, and the factorization of such amplitudes 
is well-known from multiplicative renormalization. 
The unrenormalized Green function in terms of fields
$\phi_i$ factorizes as
\beq
G_{\rm un}(p_i,\Lambda,g_0) = \prod_i \sqrt{Z_i}\Big(\frac{\Lambda}{\mu},
g(\mu)\Big) \, G_{\rm ren}(p_i, \mu, g(\mu) \Big)
\eeq
where $\Lambda$ is a UV cutoff, and $g_0$/$g(\mu)$ is the unrenormalized/renormalized
coupling. $G_{\rm un}$ does not depend on $\mu$, nor does
$G_{\rm ren}$ on $\Lambda$, so that
\beq
\mu\frac{d G_{\rm ren}}{d\mu} = -\sum_i \gamma_i\Big(g(\mu)\Big)
\eeq
where $\gamma_i = (\mu d/d\mu) \ln \sqrt{Z_i}$ can only depend
on the renormalized coupling, by separation of variables.
The solution to this evolution equation (here the
renormalization group equation), is the resummed Green function.
In this example single logarithms are resummed. We wish to
exhibit a similar paradigm for the double logarithmic, or
Sudakov, case. The connection between factorization and 
Sudakov resummation was already pointed out by Mueller, Collins
and Sen \cite{MueCollSen}. We only provide a brief description 
here of a streamlined approach \cite{CLS} to Sudakov resummation 
for various reactions, that stresses common features.

We consider cross sections near the edge of phase space,
where there is not much room for additional gluonic radiation,
and that are color singlets at lowest order (true QCD
processes can be treated in a similar fashion \cite{CLS,KS}).
Examples are the $e^+e^-$ total
cross section to hadrons near unit thrust,
deep-inelastic scattering near unit Bjorken $x$, 
Drell-Yan production near threshold, etc. Let us call
the edge of phase space in such cases the elastic limit.
In the integral over virtual and real gluonic degrees of 
freedom, which takes place in the cross section, 
the most important momentum regions
are, e.g. in Drell-Yan: (i) fast, almost collinear partons in the two
incoming jets, (ii) far off-shell, short-distance partons
that result from the collision of the two incoming jets,
and produce the off-shell vector boson, and (iii) 
soft gluons that couple the two incoming jets with momenta
$p_1$ and $p_2$.
Near the elastic limit, the cross section factorizes 
\cite{ster87} into
corresponding hard, a soft, and two jet functions that summarize these degrees
of freedom
\bea
{\tilde \sigma}(N) &=&
\int_0^\infty dw\, e^{-Nw} \sigma(w) =H(p_1/\mu,p_2/\mu,\zeta_i)\; {\tilde S}(Q/\mu N,\zeta_i) 
\nonumber \\
&& \quad\quad \times
{\tilde J}_1(p_1\cdot \zeta_1/\mu,Q/\mu N)\;  
{\tilde J}_2(p_2\cdot \zeta_2/\mu,Q/\mu N)\, ,
\label{transform}
\eea
Let us discuss the variables that occur in the above
equation. $Q$ is the hard scale of the process 
(the invariant mass of the vector boson produced
in Drell-Yan) and $w$ is a dimensionless weight function that 
is defined to vanish in the elastic limit, is
insensitive to collinear splittings of partons, and to
additional soft radiation 
and is additive near the edge of phase space, i.e.
$w = w_1 + w_2 + w_s$ where $w_i$ are contributions to 
the weight function from momenta in the jets, and $w_s$ is
the contribution
from soft momenta. In DY one can choose $w=1-z = 1-Q^2/s$.
Further, $N$ is the moment variable Laplace-conjugate to $w$, 
and is large near the elastic limit. 
Arbitrary variables, on which the physical cross section
may not depend, are most generally the factorization scale $\mu$ 
and the vectors $\zeta_i$, which are necessary
to define the jet functions $J_i$. The latter can be thought of 
as gauge-fixing vectors, as one may compute the jet functions
in different gauges, as long as the total cross section
does not depend on these gauges. It is the arbitrariness in these
vectors that allows one to extend the earlier arguments
for the single logarithm case to the Sudakov double logarithmic
one.

One proceeds by acting
on this factorized form (\ref{transform})
with {\it two} differential operators with respect to 
the arbitrary variables just mentioned:
\beq
\mu\frac{d}{d\mu}\quad\quad , \quad\quad p_i\cdot\zeta\frac{d}{dp_i\cdot\zeta}
\quad( i=1,2)\, .
\eeq
Using separation of variables, and solving the (double) differential
evolution equations obtained with the above operators, one arrives
at \cite{CLS}:
\beq
\ln {\tilde \sigma}(N)
= C^{(0)}+(\alpha_s/\pi)
\bigg [   A^{(1)}\ln^2N
+  B^{(1)}\ln N + C^{(1)}\bigg ]\, 
\label{oneloop}
\eeq
where $A^{(1)},B^{(1)},C^{(1)}$ have to be fitted from a 
next-to-leading order calculation. As expected, the resummed
cross section is an exponential, with at most double
logarithms, while higher powers of 
$\ln N$ in the exponent arise only from the expansion of the running 
coupling. In the next section we consider a specific and important
example, viz. the resummation of Sudakov
logarithms in Higgs production at the LHC.

\section{The Resummed Higgs Production Cross Section}

The presently allowed SM
Higgs mass ranges from about 70 GeV (from direct searches at
LEP \cite{MorHLL}) to about 700 GeV from unitarity/triviality constraints
\cite{HULL}.
In the MSSM the lower limits on the two scalar Higgs bosons
$h$ and $H$, and the pseudoscalar $A$ are about 60 GeV \cite{MorHLL}. 
The theoretical upper limit on 
the $h$ mass is about 130 GeV \cite{shULL}.
At the LHC the dominant production process is gluon-gluon to
Higgs via a top (and bottom, about 10\%) quark loop. Two approximations are in order. 
First, we shall neglect initial states
involving quarks (they contribute only 10\% at NLO). 
Second, we would like to consider the gluon-gluon-Higgs coupling $\kappa_\phi$
as effectively pointlike. This can be done by taking the infinite
top mass limit, supplemented by low-energy theorems \cite{lowenergy}.
Comparing the NLO infinite top mass limit result \cite{nloinfmass}
with the full analytic massive NLO result \cite{SDGZ}
one finds a difference less than 10 \% for the full range for the
SM, as well as for the MSSM provided $\tb$ is not too large.
We may then write for the $d$-dimensional partonic cross sections ($d=4-2\epsilon$)
\beq
\hat\sigma^\phi=\hat\sigma^\phi_{gg} = \sigma^\phi_0~\kappa_\phi~\rho_\phi
(z,M_\phi^2/\mu^2,\epsilon)
\label{rhodef}
\eeq
with the coefficients
\bea
\sigma^{h,H}_0 & = & g_t^{h,H}~\frac{G_F\alpha_{s,B}^2N_C C_F}{1152\sqrt{2}\pi}
\frac{\Gamma^2(1+\epsilon)}{1-\epsilon}
\left( \frac{4\pi}{m_t^2}\right)^{2\epsilon}, \\
\sigma^A_0 & = & g_t^A~\frac{G_F \alpha_{s,B}^2 N_C C_F}{512\sqrt{2}\pi}
\frac{\Gamma^2(1+\epsilon)}{1-\epsilon}
\left( \frac{4\pi}{m_t^2}\right)^{2\epsilon},
\label{sigzero}
\eea
where $\alpha_{s,B}$ is the bare strong coupling constant (with
dimension $2\epsilon$) and $g_t^\phi (\phi = h,H,A)$ denote the
modified top Yukawa couplings normalized to the SM coupling, which are
given in \cite{SDGZ}. 
The effective coupling constants $\kappa_\phi$ are given to NNLO
in \cite{KLS,Kniehl}. The correction factor may be expanded
perturbatively
\beq
\rho_\phi(z,M_\phi^2/\mu^2,\alpha(\mu^2),\epsilon) = \sum_{n=0}^{\infty}~
\alpha^n(\mu^2) \rho_\phi^{(n)}(z,M_\phi^2/\mu^2,\epsilon)
\label{sumpth}
\eeq
where we define $\alpha(\mu^2) \equiv \alpha_s(\mu^2)/\pi$.
The lowest and next order components of $\rho$ are \cite{nloinfmass}
\begin{eqnarray}
\rho_\phi^{(0)}(z,M_\phi^2/\mu^2,\epsilon) & = & \delta(1-z) \\
\rho_{h,H}^{(1)}(z,M_\phi^2/\mu^2,\epsilon) & = & 
\Big(\frac{\mu^2}{M_\phi^2}\Big)^\ep C_A\Big\{ -\frac{z^\epsilon}{\epsilon}
\left[\frac{1+z^4+(1-z)^4}
{(1-z)^{1+2\epsilon}}\right]_+ \nonumber \\
&+&\delta(1-z)\left(\frac{11}{6\epsilon}
 + \frac{203}{36} + \frac{\pi^2}{3}\right)
-\frac{11}{6} z^\epsilon (1-z)^{3-2\epsilon} \Big\}
\label{rho1} \\
\rho_A^{(1)} & = & \rho_{h,H}^{(1)} + 2 \left(\frac{\mu^2}{M_\phi^2}
\right)^\eps C_A \delta(1-z)
\label{rho1a}
\end{eqnarray}
We have implicitly redefined the scale $\mu$ by
$\mu^2 \rightarrow \mu^2\exp [-(\ln(4\pi) - \gamma_E)]$,
and scaled an overall $1/z$ into the parton
distribution functions \cite{KLS}.

Let us now construct a resummed expression for
$\rho_\phi(z,M_\phi^2/\mu^2,\alpha,\epsilon)$, or
rather for its Mellin tranform
$\tilde \rho_\phi(N,\ldots) = \int_0^1 z^{N-1}
\rho_\phi(z,\ldots)$.
We could follow the methods outlined in the first section,
but can in fact simplify further. Assuming the 
Higgs cross section factorizes near the elastic
edge of phase space, one would arrive at a resummed
cross section of the form (\ref{oneloop}). By acting
on this equation with $(M_\phi^2 d/dM_\phi^2)$ one trivially obtains
an evolution equation in the Higgs mass of the form:
\beq
M_\phi^2 \frac{d}{dM_\phi^2} \tilde \rho_\phi(N,M_\phi^2/\mu^2,\alpha(\mu^2),\epsilon)\! = \!
  \tilde W_\phi(N,M_\phi^2/\mu^2,\alpha(\mu^2),\epsilon) 
\tilde \rho_\phi(N,M_\phi^2/\mu^2,\alpha(\mu^2),\epsilon)\, .
\label{sudevz}
\eeq
In order to solve eq.~(\ref{sudevz}) we must impose a boundary condition,
and find the evolution kernel.
We may in fact use \cite{CLS,KLS} the boundary condition 
$\tilde \rho_\phi(N,M_\phi^2/\mu^2=0,\alpha(\mu^2),\epsilon)=1$ for the moments,
or in $z$-space $\rho_\phi(z,M_\phi^2/\mu^2=0,\alpha(\mu^2),\epsilon) = \delta(1-z)$.
The solution to eq.~(\ref{sudevz}) is then
\beq
\tilde \rho_\phi(N,M_\phi^2/\mu^2,\alpha(\mu^2),\ep) 
= \exp\left[ \int_0^{M_\phi^2}
 {d \xi^2 \over \xi^2} 
{\tilde W}_\phi \biggl(N,{\xi^2\over \mu^2},\alpha(\mu^2),\ep\biggr) 
\right]\, ,
\label{sudsolveN}
\eeq
which may be expanded, using renormalization group invariance, as
\beq
{\tilde \rho}_\phi(N,\frac{M_\phi^2}{\mu^2},\alpha(\mu^2),\epsilon) =  
\exp\biggl[\int_0^1dzz^{N-1}\int_0^{\nu \frac{M_\phi^2}{\mu^2}}\frac{d\lambda}{\lambda}
\Big\{\alpha(\lambda,\alpha(\mu^2),\epsilon)W_\phi^{(1)}(z,1/\nu,\epsilon) +\ldots
\Big\}\biggr]
\label{eqnsoln}
\eeq
with $\nu(z)$ an arbitrary function. 
The one loop coefficient of the evolution kernel $W_\phi$
can then be derived from a low order calculation of the correction factors
$\rho_\phi$
via
\beq
{\tilde W}_\phi^{(1)}(N,1,\epsilon)=
(M_\phi^2/\mu^2)^{\epsilon}\, 
M_\phi^2 \frac{\partial }{ \partial M_\phi^2}\, 
{\tilde \rho}_\phi^{(1)}(N,M_\phi^2/\mu^2,\epsilon)\ ,
\eeq
The general structure of the result is, in $z$-space
\beq
W_\phi^{(1)}(z,1,\ep)=\delta(1-z)f_\phi^{(1)}(\ep)
+z^\ep\left({g^{(1)}(z,\ep)\over (1-z)^{1+2\ep}}\right)_+
+h^{(1)}(z,\ep)\ ,
\label{wdecomp}
\eeq
where the coefficient functions $f_\phi^{(1)},\ g^{(1)},
\ h^{(1)}$
are regular functions of their arguments at $z=1$. 
We drop the term $h^{(1)}$ as it is of order $1/N^4$ in moment space.
After rescaling to incorporate the factor $(1-z)^{-2\ep}$ and combining the plus
distribution with the Mellin transform in eq.~(\ref{eqnsoln}), we see
that the relevant function to approximate is
$(z^{N-1}-1)g^{(1)}(z,\ep)$. We approximate the residue function
$g^{(1)}(z,\ep)$ in three schemes, 
which are defined by
\bea
\mbox{scheme~} \alpha &:& \frac{1}{C_A}
(z^{N-1}-1)g^{(1)}(z,\ep)\go (z^{N-1}-1)~2
\nonumber\\
\mbox{scheme~} \beta &:& \frac{1}{C_A}
(z^{N-1}-1)g^{(1)}(z,\ep) \go (z^{N-1}-1)~2 
   - (1-z)(2z^2-4z-2z^3) \nonumber\\
\mbox{scheme~} \gamma &:& \frac{1}{C_A}
(z^{N-1}-1)g^{(1)}(z,\ep) \go (z^{N-1}-1)~2
   - (1-z)(2z^2-4z-2z^3) \nonumber \\
& & \hspace*{7.0cm} - 4 z^{N-1} (1-z)\,. 
\label{schemes}
\eea
The minimal scheme $\alpha$ involves replacing $g^{(1)}(z,\ep)$ simply
by $g^{(1)}(1,\ep)$, scheme $\beta$ includes all terms of ${\cal
  O}(1)$ in the exponent, whereas scheme $\gamma$ includes in addition
some $O(\ln^i N/N)$ terms.  Using the one-loop
evolution kernel $W_\phi^{(1)}$ we can now construct the resummed
expressions for the Higgs production correction factor in the three
schemes. Although these expressions are still divergent for $\ep\go
0$, the divergences may be cancelled by mass factorization and
renormalization for which we chose the $\msb$ scheme.

The final results for the resummed cross sections in moment
space are given in \cite{KLS}, both for Higgs production
and the Drell-Yan process. These two processes are very 
similar from the soft gluon point of view, the main
difference being that Higgs production is driven by
gluon fusion, and Drell-Yan by quark-antiquark annihilation.
Rather than evaluate the resummed answers numerically - which
involves the difficult problem of treating the infrared
renormalon - we expanded the resummed answers \cite{KLS} for Higgs 
production and Drell-Yan\cite{DYres} to NNLO in QCD
perturbation theory, in the above three schemes. The
answers are expressed in terms of plus-distributions
${\cal D}_i(z)$  and logarithms ${\cal E}_i(z)$ (which are integrable
but large near the edge of phase space)
\beq
{\cal D}_i(z) = \left[\frac{\ln^i(1-z)}{1-z}\right]_+
\quad, \quad
{\cal E}_i(z) = \ln^i(1-z),
\eeq
and constants. Scheme $\gamma$ incorporates
the ${\cal E}$ logarithms, which have not been included in resummed cross sections 
before. At any order, the leading ${\cal E}$ logarithms and those subleading ones that
are related to the QCD running coupling are universal, as they
arise directly from the splitting function. 
They occur in exact NNLO calculations for DY and DIS \cite{NNLOex},
and we checked analytically that the scheme $\gamma$ resummed 
answer for these processes, expanded to NNLO, reproduces them.
\begin{figure}[htbp]
\vspace*{-1.1cm}
\hspace*{0cm}
\begin{center}
\begin{tabular}{cc}
\epsfig{file=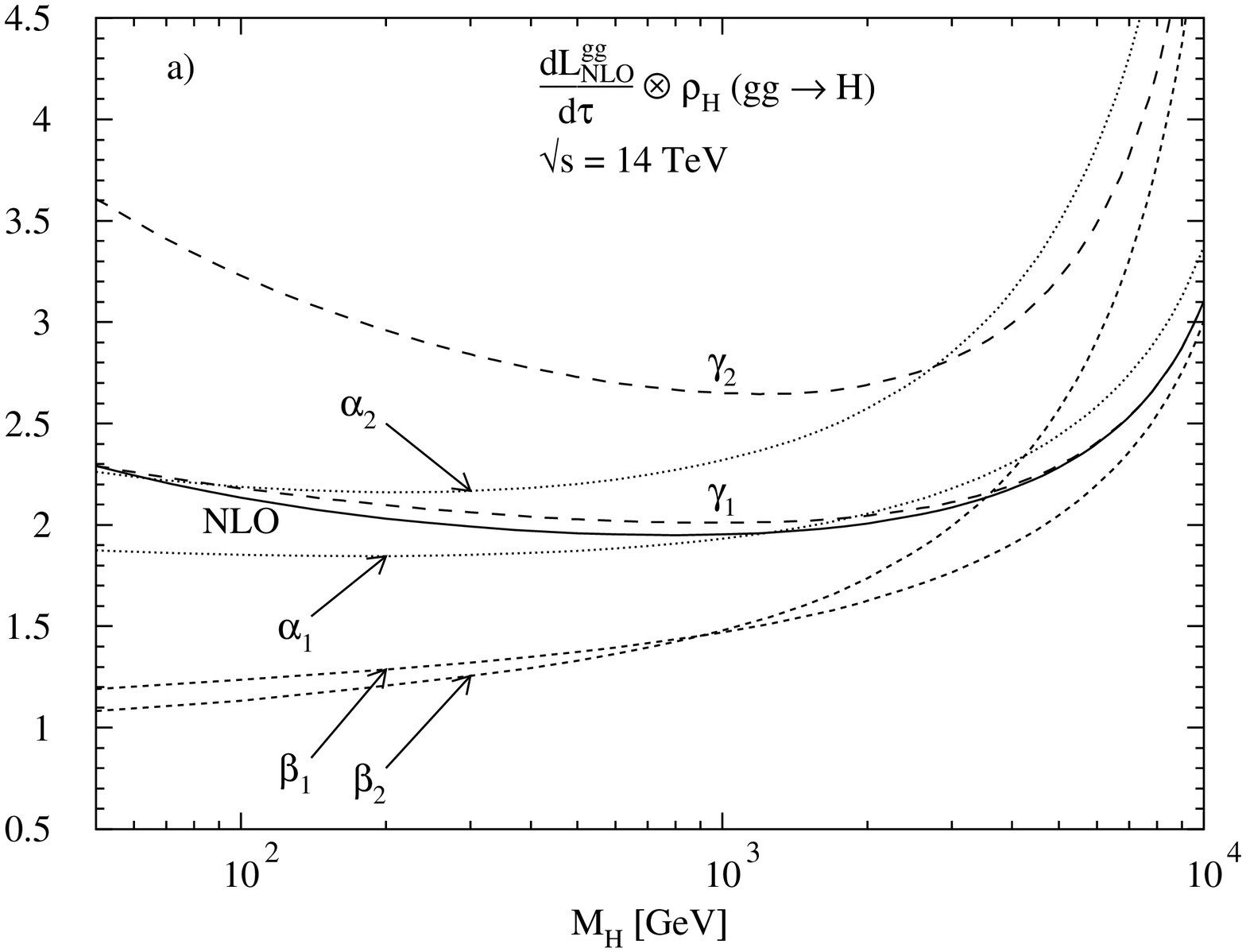,bbllx=0pt,bblly=+150pt,bburx=575pt,bbury=880pt,width=6.5cm,angle=-90}
\epsfig{file=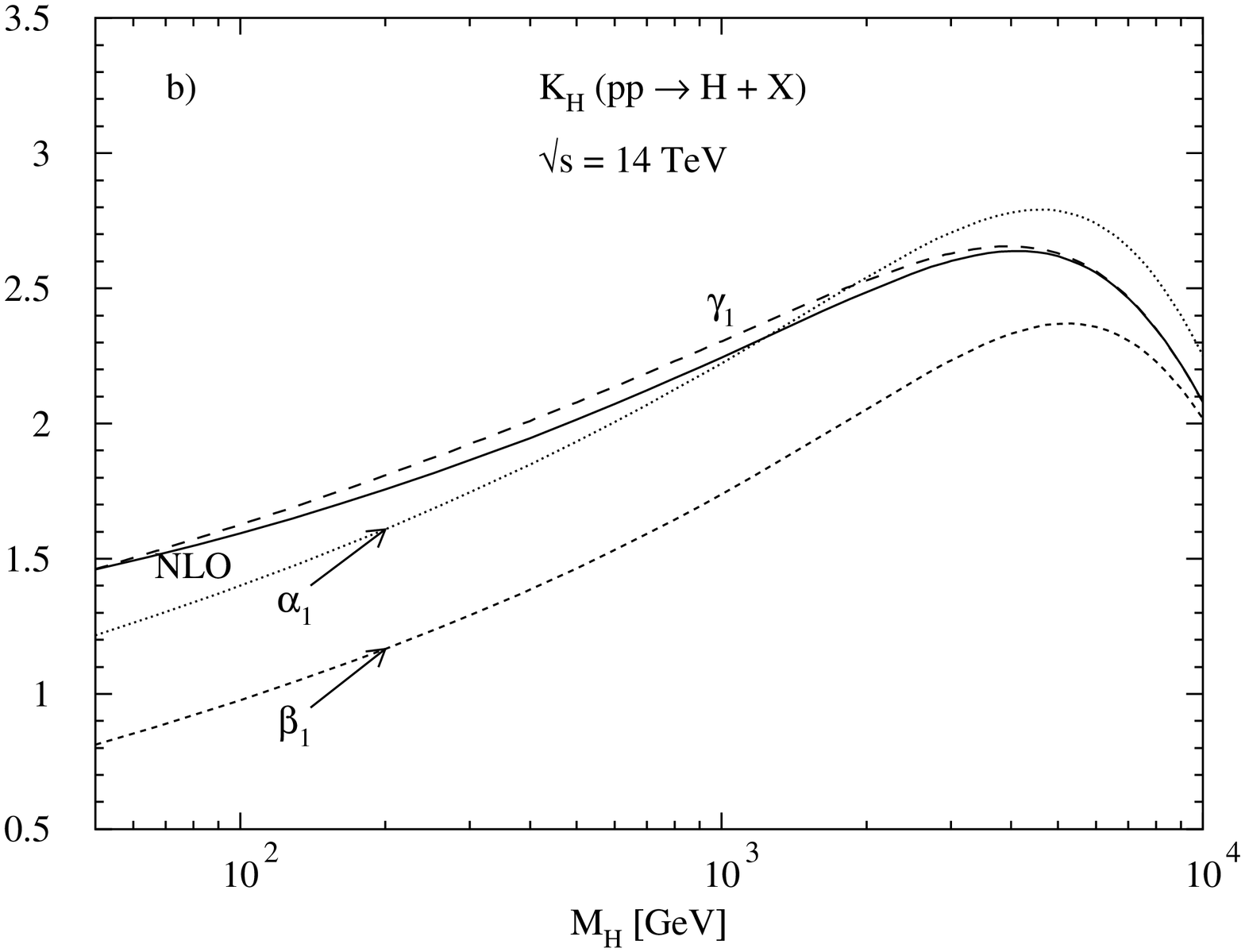,bbllx=0pt,bblly=+150pt,bburx=575pt,bbury=880pt,width=6.5cm,angle=-90}
\end{tabular}
\end{center}
\caption[]{\it a) Exact (solid line) and approximate two- ($\alpha_1, \beta_1, \gamma_1$)
and three-loop  ($\alpha_2, \beta_2, \gamma_2$) partonic
K-factors, convoluted with the NLO gluon-gluon luminosity $d{\cal
 L}^{gg}_{NLO}/d\tau$, where $\tau = M_\phi^2/s$, 
in the heavy top-mass limit and in 
three different schemes, versus the
scalar Higgs mass $M_H$. We used NLO CTEQ4M parton densities
\cite{CTEQ4} and $\alpha_s$ ($\Lambda_{\msb}^{(5)} = 202$ MeV).  b)
Hadronic NLO K-factor using LO CTEQ4L parton densities \cite{CTEQ4}
and $\alpha_s$ ($\Lambda_{\rm LO}^{(5)} = 181$ MeV) for the LO cross
section and including the NLO contributions from $\kappa_H$.}
\label{Kf1}
\end{figure}
In Fig.~\ref{Kf1} we present
the correction factors for SM Higgs production at the LHC, which
coincide with the correction factors of MSSM scalar Higgs boson
production for small \tb. For MSSM pseudoscalar
Higgs production we show similar results in \cite{KLS}.  
In Fig.~\ref{Kf1}a the ``partonic'' K-factors,
obtained from folding the correction factors $\rho_\phi$ with NLO
parton densities and using a NLO strong coupling for all orders of the
cross sections, are presented. For comparison we show in
Fig.~\ref{Kf1}b the corresponding NLO ``hadronic''
K-factors 
normalized to the LO cross sections evolved with LO parton densities
and $\alpha_s$.  Whereas the former indicate the rate of convergence
of the individual-order contributions within a fixed order calculation, the 
latter exhibit the convergence of the perturbative approach to the physical
(hadronic) quantities.  We observe from Fig.~\ref{Kf1} 
that at NLO scheme $\gamma$, remarkably, reproduces the exact NLO calculation
almost exactly for the full range of the SM Higgs mass $M_H \gsim 70$
GeV (similar results are obtained for the neutral Higgses in the MSSM)
schemes $\alpha$ and $\beta$ agree with the exact result only for
$M \gg 1$ TeV (the agreement of scheme $\alpha$ in the
intermediate Higgs mass range is accidental).  Moreover, note that the
NNLO corrections to the partonic cross sections in scheme $\gamma$ are
still very significant.  
We checked \cite{KLS} in Drell-Yan that similarly good agreement is obtained
in scheme $\gamma$ in NLO and NNLO for the same kinematics \cite{NNLOex}.
Keep in mind that full NNLO predictions for hadronic cross sections
require NNLO parton densities, which are not yet available. 
Should the size of the NNLO corrections to the physical cross
section warrant concern about the convergence of the perturbative
approach, our resummed answer provides a tool to control such
large corrections.

\section*{Acknowledgments}

I thank Harry Contopanagos, Michael Kr\"{a}mer, Michael Spira and George
Sterman for a very pleasant collaboration.

\end{document}